\documentclass[groupedaddress]{article}
\usepackage[top=32mm, bottom=30mm, left=25mm, right=25mm]{geometry}
\usepackage{amsmath,amsfonts,amssymb,color,bbm}
\usepackage{amsmath,amssymb,bm}
\usepackage{amsfonts}
\usepackage{amscd}
\usepackage{delarray}
\usepackage{graphicx}
\usepackage{soul}
\newcommand{\rw}[1]{\textcolor{black}{#1}}

\begin{document}


\title{Non-classical properties of the e.m. near field of an atom in spontaneous light emission.}


\author{ Vincent Debierre, $^{[1]}$, Brian Stout$^{[1]}$, Thomas Durt$^{[1]}$ \\ $^{[1]}$Aix-Marseille Univ, CNRS, Centrale Marseille, Institut Fresnel, 13397 Marseille, France. }







\maketitle
\begin{abstract}We use Glauber's correlation function as well as the Green functions formalism to investigate, in the case of a dipolar atomic transition, the causal behaviour of the spontaneously emitted electromagnetic field, in the A.p coupling.  This brings us to examine the role played by the longitudinal electric field, which is not described in terms of photonic (transverse) degrees of freedom.
\end{abstract}

keywords : Spontaneous emission, causality, photon wave function, Green function

\section{Introduction}\label{sec1}

The question of causality in the spontaneous emission of light by an excited atomic electron has a long history. In his pioneering work, Fermi concluded \cite{FermiCausal} that, within the framework of the Wigner-Weisskopf approximation (exponential decay), the emitted electric field vanishes outside the light cone centred at the position of the emitting atom and the time at which the emission starts (hereafter simply referred to as `the light cone'). This result has been amended by Hegerfeldt, who noticed \cite{HegerfeldtFermi} that the absence of negative frequencies from the emitted field prevents causal propagation. 
Increased interest in the total absorption of a light wave by an atom \cite{LeuchsFourPi,LeuchsFresnel,LeuchsMoving,Stout} shines the spotlight on this old problem, since perfect absorption is the time-reversed process of spontaneous emission in a vacuum. 

Recently, spontaneous light emission in the mid- and far-field regions of emission was revisited \cite{CausalMidFar}, confirming Hegerfeldt's objection to Fermi's causal result. We also confirmed Shirokov's intuition \cite{Vlad} that, since the electronic wave functions are of finite extent, the light cone is blurred on a length scale of the order of the Bohr radius, even if one overlooks Hegerfeldt's objections. To avoid confusion between features that are only apparently non-causal, and features that are indeed non-causal, it is preferable to isolate the Green's function of the problem and study its properties, independently from the specific details of the atomic light source. In so doing, we will establish a clear parallel between the quantum and classical treatments of light emission and highlight the crucial role of the longitudinal component of the emitted electric field for causality, which sheds a new light on standard photodetection theory \cite{Titulaer,GlauberHouches}. 

Description of light detection and the theory of optical coherence \cite{GlauberKoh} have motivated and justified the use of the so-called Glauber first order correlation function, which can alternatively be interpreted as a single photon wave function \cite{Foxmulder}, and/or as a single photon electric field \cite{HarocheRaimond}. 

However, most attention so far has been devoted to the wave functions of free photons in the absence of sources \cite{RaymerSmith}. Notable exceptions, besides our earlier work \cite{CausalMidFar}, are contributions by Sipe \cite{Sipe} and Scully and Zubairy \cite{Foxmulder}, wherein it is shown how the causal propagation of spontaneously emitted photons can be established through the photon wave function formalism. This (standard) result is not fully comfortable however because, in those approaches, the quantum Green function that is derived in the absence of the longitudinal field is essentially equivalent to its classical counterpart, which results from the joint contribution of transverse and longitudinal fields. This would suggest that the classical-quantum correspondence is not fully verified, a rather suspicious hypothesis, which brought us to study the quantum Green function corresponding to the A.p coupling between the electron and the Maxwell fields and not the E.r coupling as has been done before \cite{Sipe,Foxmulder}. In this work, we show that then the picture is more complicated. Our main results are the following: 

1) if we resort to the A.p interaction Hamiltonian, there appears at the level of the photon wave function a new source of non-locality, different from the two aforementioned (Hegerfeldt and Shirokov) types of non-locality (which essentially amount to a blurring of the causal light cone \cite{CausalMidFar,VincePhd22}). Disregarding this blurring effect, the mid and far field are very close to those computed, in the standard approach, with the E.r interaction Hamiltonian, at the dipolar approximation. This is not so however in the case of the near field where a new type of non-locality appears (section \ref{2}). 

2)  A possible way to get rid of the problem consists of pursuing the analogy with what happens in classical electrodynamics, where this non-locality is compensated by the instantaneous longitudinal (Coulomb) field generated by the quantum dipole \cite{Jackson,Rahmani}. If we follow this strategy however, we predict that, even after the emission of the photon, when the atom has returned to the ground state, the near field surrounding the atom remains disturbed by the presence of a remnant static quantum field. This is the price to pay to restore causality in this approach (section \ref{3}).

 Another strategy for tackling our problem consists in considering the longitudinal photon wave function (section \ref{4}). The result obtained in this approach does not suffer from the aforementioned memory effect derived in section 3, pursuing the analogy with classical Green functions. It predicts however the appearance of non-local effects, which remain confined to the near field.

\section{The decay of a two-level atom: dipolar and multipolar contributions to the near-field.} \label{2}
\subsection{The decay of a two-level atom} \label{sec:AtomQED}

In the rotating wave approximation, which is well justified \cite{EdouardIsa} in the case of spontaneous emission in the vacuum, 
the state vector of the atom-field system at time $t\geq0$ reads,
\begin{align}\begin{split} 
          \mid\!\psi\left(t\right)\rangle&=c_{\mathrm{e}}\left(t\right)\mathrm{e}^{-\mathrm{i}\omega_{\mathrm{e}}t}\mid\!\mathrm{e},0\rangle \\ &\qquad+\sum_{\lambda=\pm}\int \mathrm{d}^3k\,c_{\mathrm{g},\lambda}\left(\mathbf{k},t\right)\mathrm{e}^{-\mathrm{i}\left(\omega_{\mathrm{g}}+ c| | \mathbf{k}|| \right) t}\mid\!\mathrm{g},1_{\lambda,\mathbf{k}}\rangle \;, \label{eq:Ansatz}
 \end{split}\end{align}
where $\mid\!\mathrm{e},0\rangle$ means that the atom is in its excited state and the field contains no photons and $\mid\!\mathrm{g},1_{\lambda,\mathbf{k}}\rangle$ means that the atom is in its ground state and the field contains a photon of wave vector $\mathbf{k}$ and polarisation $\lambda$.\\ We assume that  the atom is rapidly excited at time $t=0$ so that, for $t<0$, $c_{\mathrm{e}}\left(t\right)=0$ while $c_{\mathrm{e}}\left(t=0\right)=1$, and $\forall t\leq 0, \forall\mathbf{k}\in\mathbb{R}^3\forall\lambda=\pm,\,c_{\mathrm{g},\lambda}\left(\mathbf{k},t\right)=0$.

The equations of motion are given by
\begin{align}\begin{split}
&\dot{c}_{\mathrm{e}}\left(t\right)=-\frac{\mathrm{i}}{\hbar}\sum_{\lambda=\pm}\int \mathrm{d}^3\mathbf{k} G_\lambda^*\left(\mathbf{k}\right)c_{\mathrm{g},\lambda}\left(\mathbf{k},t\right)\mathrm{e}^{-\mathrm{i}\left(c||\mathbf{k}||-\omega_{\mathrm{e}}+\omega_{\mathrm{g}}\right)t}, \\
&\dot{c}_{\mathrm{g},\lambda}\left(\mathbf{k},t\right)=-\frac{\mathrm{i}}{\hbar }G_\lambda\left(\mathbf{k}\right)c_{\mathrm{e}}\left(t\right)\mathrm{e}^{\mathrm{i}\left(c||\mathbf{k}||-\omega_{\mathrm{e}}+\omega_{\mathrm{g}}\right)t} \;.\label{eq:ExactEQg}
\end{split}\end{align}
\rw{Let us for the time being focus on the 1S-2P transition in hydrogen atom.}
In standard approaches \cite{Foxmulder} the coupling factors $G_\lambda\left(\mathbf{k}\right)=\langle1\mathrm{s},1_{\lambda,\mathbf{k}}\mid\!\hat{H}_I\!\mid\!2\mathrm{p}\,m_2,0\rangle$ are estimated in the dipolar approximation for the E.r coupling, in which case we find 

\begin{eqnarray} \label{eq:MatrixDipole}
\langle1\mathrm{s},1_{\lambda,\mathbf{k}}\mid\!\hat{H}^{E.r-dip.}_I\!\mid\!2\mathrm{p},0\rangle
=\mathrm{-i}\sqrt{\frac{\hbar c  k}{16\pi^3\epsilon_0 }}\bm{\epsilon}_{\left(\lambda\right)}^*
\left(\mathbf{k}\right)\cdot\bm{\mu}_{g-e.} \;,
\end{eqnarray}
where we  denoted $\bm{\epsilon}_{\left(\lambda\right)}$ the direction of the polarization of the e.m. mode, and we defined the quantum dipole associated to the transition through:
\begin{equation} \label{dipoleQ}\bm{\mu}_{g-e.}=\langle1\mathrm{s}\mid\! e\bm{r}\! \mid\!2 \mathrm{p}\rangle \;. \end{equation}

Due to degeneracy of the $2P$ level, a pure $2P$ state is combination of the three $L_Z$ eigenstates  $\mid\! 2\mathrm{p},m_2=-1\rangle$, $ \mid\!2\mathrm{p},m_2=0\rangle$ and $\mid\! 2\mathrm{p},m_2=+1\rangle$ and it can be shown \cite{Joachain} that $\bm{\mu}_{g-e.}=\sqrt 2(2^7/3^5)ea_0 \bm{\xi}_{m_2}$ with $a_0$ the Bohr radius, where the $\bm{\xi}_{m_{2}}$ are given by
\begin{subequations} \label{eq:Xi}
\begin{eqnarray}
\bm{\xi}_0&=\hat{z},\\
\bm{\xi}_{\pm1}&=\mp\frac{\hat{x}\pm\mathrm{i}\hat{y}}{\sqrt{2}}.
\end{eqnarray}
\end{subequations}

In the A.p coupling, the coupling $G_\lambda\left(\mathbf{k}\right)=\langle1\mathrm{s},1_{\lambda,\mathbf{k}}\mid\!\hat{H}_I\!\mid\!2\mathrm{p}\,m_2,0\rangle$
can be computed exactly, taking all multipolar contributions into account \cite{FacchiPhD2} (see also Ref. \cite{VincePhd22}, section 6.1.4); it reads, if the 2P state is one of the three states defined through (\ref{eq:Xi}) \begin{align}\begin{split} \label{eq:MatrixMultipole}
 & \langle 1\mathrm{s},1_{\lambda,\mathbf{k}}\mid\!\hat{H}^{A.p-multipolar}_I\!\mid\!2\mathrm{p}\,m_2,0\rangle  =\\ & \quad \langle 1\mathrm{s}\mid\!\sqrt{\frac{\hbar}{16\pi^3\epsilon_0 ck}}e^{-i\mathbf{k}\cdot 
\mathbf{r}}\bm{\epsilon}_{\left(\lambda\right)}^{\ast}\left(\mathbf{k}\right)\cdot \frac{e\hbar\mathrm{i}\mathbf{\nabla}}{m}\mid\!2 \mathrm{p}m_2\rangle =\mathrm{-i}\omega_0\sqrt{\frac{\hbar}{16\pi^3\epsilon_0 c k}}\frac{2^{\frac{15}{2}}}{3^5}e a_0\frac{\bm{\epsilon}_{\left(\lambda\right)}^*\left(\mathbf{k}\right) \cdot\bm{\xi}_{m_2}}{\left[1+\left(\frac{2}{3}a_0||\mathbf{k}||\right)^2\right]^2},
\end{split}\end{align} 
where $\omega_0$ represents the Bohr frequency ($\omega_0=\omega_e-\omega_g$).

Close to resonance, $ck\approx \omega_0$, and $1+(\frac{2}{3}a_0||\mathbf{k}|| )^2 \approx 1,$ because optical wavelengths are much larger than the Bohr radius. Comparing  with the expression (\ref{eq:MatrixDipole}), we get thus    
 \begin{eqnarray} \frac{\langle 1\mathrm{s},1_{\lambda,\mathbf{k}}\mid\!\hat{H}^{E.r-dip.}_I\!\mid\!2\mathrm{p}\,m_2,0\rangle}{\langle 1\mathrm{s},1_{\lambda,\mathbf{k}}\mid\!\hat{H}^{A.p-multipolar}_I\!\mid\!2\mathrm{p}\,m_2,0\rangle}\approx \frac{ck}{\omega_0}\approx 1.\end{eqnarray}

The E.r coupling factor is thus close to its A.p counterpart at resonance. This implies among others that the estimate of the lifetime of the excited state, when it is performed making use of the Fermi golden rule, is the same in the A.p and in the E.r coupling, where only the expression of the coupling $G_\lambda\left(\mathbf{k}\right)$ at resonance plays a role. The equality does not hold for high frequencies however, which has far reaching consequences as we shall show.
 \subsection{Glauber first order correlation  function and quantum Green function.}
 The Glauber first order correlation  function is written \cite{GlauberKoh,Foxmulder}
\begin{equation} \label{eq:GeneralWaveFunction}
\begin{aligned} [b]
\bm{\psi}_\perp\left(\mathbf{x},t\right)&\equiv\langle\mathrm{g},0\!\mid\hat{\mathbf{E}}_\perp\left(\mathbf{x},0\right)\mid\psi\left(t\right)\rangle
\end{aligned}
\end{equation}
where $\hat{\mathbf{E}}_\perp$ represents the transverse part of the electric field operator defined in Supplementary Material (section 2).

As is well-established in the standard theory of photodetection \cite{GlauberKoh,Foxmulder,HarocheRaimond}, the modulus squared of $\bm{\psi}_\perp\left(\mathbf{x},t\right)$  in the expression (\ref{eq:GeneralWaveFunction}) is proportional to the probability of detecting a photon at time $t$ in a detector located at position $\mathbf{x}$ and can be interpreted as a kind of photon wave function\footnote{The interpretation is unambiguous in the single photon case, as discussed in Ref.  \cite{Aspect}, which is precisely the case that we are dealing with here.}. In analogy with the classical Poynting density, the photon wave function $\bm{\psi}_\perp\left(\mathbf{x},t\right)$ can also be interpreted as the single photon electric field generated by the quantum dipole. In the rest of the paper we shall thus, depending on the context, sometimes call $\bm{\psi}_\perp$ the photon wave function, and sometimes the single photon (transverse) electric field, keeping in mind that it is a quantum quantity, not to be confused with the average electric field for instance, which is equal to zero in the case of a single photon state. It also differs from the classical electric field by the fact that it is a complex-valued field.

\subsubsection{Arbitrary initial polarisation (A.p coupling).}
After integrating (\ref{eq:ExactEQg}) over time to find $c_{\mathrm{g},\lambda}\left(\mathbf{k},t\right)$, one gets, making use of (\ref{eq:GeneralWaveFunction}), an expression of the photon wave function (or single photon electric field) $\bm{\psi}_\perp\left(\mathbf{x},t\right)$ in function of the survival amplitude of the excited atomic state $c_e(t')$ \cite{Foxmulder}.
 
 The expression  of  $\bm{\psi}_\perp\left(\mathbf{x},t\right)$ has been  established  in the past for the A.p coupling \cite{VincePhd22,CausalMidFar}. 
 We established for instance in \cite{CausalMidFar} (see Supplementary Material section 2 for a derivation ``from scratch'') that the exact photon wave function is given by 
\begin{align}
&\bm{\psi}^{A.p-multipolar}_\perp\left(\mathbf{x},t\right)\label{eq:PsiBeta}\\\nonumber =&-\mathrm{i}\frac{2^{\frac{7}{2}}}{3^4}\frac{\hbar e}{\epsilon_0m_ea_0}\int_0^{+\infty}\frac{\mathrm{d}k}{\left(2\pi\right)^2}k^2\frac{\mathrm{e}^{-\mathrm{i}ckt}}{\left[1+\left(\frac{k}{k_\mathrm{X}}\right)^2\right]^2}\mathbf{I}\left(k,||\mathbf{x}||\right)\int_0^t\mathrm{d}t'\,c_{\mathrm{e}}\left(t'\right)\mathrm{e}^{\mathrm{i}\left(ck-\omega_0\right)t'}, 
\end{align}
with $k_X=3/2a_0$, while the vector $\mathbf{I}$ is defined through

\begin{eqnarray} 
&I^{\left(x,y\right)}\left(k,||\mathbf{x}||\right)=\mathrm{i}\frac{\xi_{m_2}^{\left(x,y\right)}}{k||\mathbf{x}||}\left[\vphantom{\frac{1}{\left(k||\mathbf{x}||\right)^2}}\left(\mathrm{e}^{-\mathrm{i}k||\mathbf{x}||}-\mathrm{e}^{\mathrm{i}k||\mathbf{x}||}\right)-\frac{\mathrm{i}}{k||\mathbf{x}||}\left(\mathrm{e}^{-\mathrm{i}k||\mathbf{x}||}+\mathrm{e}^{\mathrm{i}k||\mathbf{x}||}\right)-\frac{1}{\left(k||\mathbf{x}||\right)^2}\left(\mathrm{e}^{-\mathrm{i}k||\mathbf{x}||}-\mathrm{e}^{\mathrm{i}k||\mathbf{x}||}\right)\right],\nonumber\\ &I^{\left(z\right)}\left(k||\mathbf{x}||\right)=-2\mathrm{i}\frac{\xi_{m_2}^{\left(z\right)}}{k||\mathbf{x}||}\left[\vphantom{\frac{1}{\left(k||\mathbf{x}||\right)^2}}-\frac{\mathrm{i}}{k||\mathbf{x}||}\left(\mathrm{e}^{-\mathrm{i}k||\mathbf{x}||}+\mathrm{e}^{\mathrm{i}k ||\mathbf{x}||}\right)-\frac{1}{\left(k||\mathbf{x}||\right)^2}\left(\mathrm{e}^{-\mathrm{i}k||\mathbf{x}||}-\mathrm{e}^{\mathrm{i}k||\mathbf{x}||}\right)\right]\label{eq:DipoleGreen}
\end{eqnarray}

Here we chose a coordinate system for which $\mathbf{x}$ points along the third axis $\hat{z}$. As we can see, the photon wave function contains contributions proportional to $1/||\mathbf{x}||$, $1/||\mathbf{x}||^2$ and $1/||\mathbf{x}||^3$, which are respectively known as the far-field, mid-field and near-field contributions. To first order in perturbation theory around $t=0$, we have $c_{\mathrm{e}}\left(t\right)=1$, whence 
\begin{equation} \label{eq:WWInt}
\begin{aligned} [b]
\int_0^t\mathrm{d}t'\,c_{\mathrm{e}}\left(t'\right)\mathrm{e}^{\mathrm{i}\left(ck-\omega_0\right)t'}=\int_0^t\mathrm{d}t'\,\mathrm{e}^{\mathrm{i}\left(ck-\omega_0\right)t'}
=\left[\frac{\mathrm{e}^{\mathrm{i}\left(ck-\omega_0\right)t'}}{\mathrm{i}\left(ck-\omega_0\right)}\right]_0^t
=\mathrm{i}\,\frac{1-\mathrm{e}^{\mathrm{i}\left(ck-\omega_0\right)t}}{ck-\omega_0}\; .
\end{aligned}
\end{equation}
For longer times, the Wigner-Weisskopf approximation is fully justified, in which case we have
\begin{align}\begin{split} \label{WW}\int_0^t\mathrm{d}t'\,c_{\mathrm{e}}\left(t'\right)\mathrm{e}^{\mathrm{i}\left(ck-\omega_0\right)t'}=\int_0^t\mathrm{d}t'\,\mathrm{e}^{\mathrm{i}\left(ck-(\omega_0-i\Gamma/2)\right)t'}\\ =\mathrm{i}\,\frac{1-\mathrm{e}^{\mathrm{i}\left(ck-(\omega_0-i\Gamma/2)\right)t}}{ck-(\omega_0-i\Gamma/2)}=\mathrm{i}\,\frac{1-\mathrm{e}^{\mathrm{i}\left(ck-\Omega_0\right)t}}{ck-\Omega_0}\;,
\end{split}\end{align} where we defined $\Omega_0=\omega_0-i\Gamma/2$.
With (\ref{eq:WWInt},\ref{WW}) in mind, we rewrite (\ref{eq:PsiBeta}) as
\begin{multline} \label{eq:BacktoStart}
\bm{\psi}^{A.p-multipolar}_\perp\left(\mathbf{x},t\right)=-\mathrm{i}\frac{2^{\frac{7}{2}}}{3^4}\frac{\hbar e}{\epsilon_0m_ea_0}\frac{\mathrm{e}^{-\mathrm{i}\Omega_0t}}{\left(2\pi\right)^2}\\
\left[
\begin{array}{lrclclcl}
\frac{\xi_{m_2}^{\left(x,y\right)}}{||\mathbf{x}||}&\left[\vphantom{\left(H_0^{\left(-\right)}\right)}\right.&\left.\left[H_1^{\left(-\right)}-H_1^{\left(+\right)}\right]\left(||\mathbf{x}||,t\right)\right.&-&\left.\frac{\mathrm{i}}{||\mathbf{x}||}\left[H_2^{\left(-\right)}+H_2^{\left(+\right)}\right]\left(||\mathbf{x}||,t\right)\right.&-&\left.\frac{1}{\mathbf{x}^2}\left[H_3^{\left(-\right)}-H_3^{\left(+\right)}\right]\left(||\mathbf{x}||,t\right)\right]\\
2\frac{\xi_{m_2}^{\left(z\right)}}{||\mathbf{x}||}&\left[\vphantom{\left(H_0^{\left(-\right)}\right)}\right.&&&\left.\frac{\mathrm{i}}{||\mathbf{x}||}\left[H_2^{\left(-\right)}+H_2^{\left(+\right)}\right]\left(||\mathbf{x}||,t\right)\right.&+&\left.\frac{1}{\mathbf{x}^2}\left[H_3^{\left(-\right)}-H_3^{\left(+\right)}\right]\left(||\mathbf{x}||,t\right)\right]
\end{array}
\right]\nonumber
\end{multline}
where
\begin{equation} \label{eq:ThisIsK}
\begin{aligned} [b]
H_n^{\left(\pm\right)}\left(||\mathbf{x}||,t\right)&\equiv\int_0^{+\infty}\mathrm{d}k\,\frac{k^{2-n}}{\left[1+\left(\frac{k}{k_{\mathrm{X}}}\right)^2\right]^2}\mathrm{e}^{\pm\mathrm{i}k||\mathbf{x}||}\,\frac{1-\mathrm{e}^{-\mathrm{i}\left(ck-\Omega_0\right)t}}{ck-\Omega_0},
\end{aligned}
\end{equation}
The integration goes over positive values of $k$ only, due to the fact that electro-magnetic modes are characterized by positive frequencies.
In order to compare the single photon electric field generated by the quantum dipole with its classical counterpart it will be useful to reformulate it, in a frame independent way, which gives 
\begin{align} \begin{split}\label{eq:PsiBeta2}
&\bm{\psi}^{A.p-multipolar}_\perp\left(\mathbf{x},t\right) =\\  &\qquad -\mathrm{i} \frac{c \omega_0 \left\|\bm{\mu}_{g-e.}\right\|}{8\pi^2\epsilon_0} \int_0^{+\infty}\mathrm{d}kk^2\frac{\mathrm{e}^{-\mathrm{i}ckt}}{ \left[1+\left(\frac{2}{3}a_0||\mathbf{k}||\right)^{2}\right]^2} \mathbf{\overleftrightarrow{M}}\left(k,\mathbf{x}\right)\int_0^t\mathrm{d}t'\,c_{\mathrm{e}}\left(t'\right)\mathrm{e}^{\mathrm{i}\left(ck-\omega_0\right)t'} \widehat{\bm{\mu}}_{g-e.},
\end{split}\end{align}
where $\bm{\mu}_{g-e.} \equiv \left\|\bm{\mu}_{g-e.}\right\|  \widehat{\bm{\mu}}_{g-e.}$, while the 3x3 matrix $ \mathbf{\overleftrightarrow{M}}$ is defined through  
 
\begin{align}\begin{split} \label{eq:DipoleGreenbis}
 \mathbf{\overleftrightarrow{M}}\left(k,\mathbf{x}\right)& = \frac{\mathrm{i} }{k||\mathbf{x}||}
\left[ ({\mathbb{I}}-\widehat{\boldsymbol{x}} \widehat{\boldsymbol{x}})\vphantom{\frac{1}{\left(k||\mathbf{x}||\right)^2}}\left(\mathrm{e}^{-\mathrm{i}k||\mathbf{x}||}-\mathrm{e}^{\mathrm{i}k||\mathbf{x}||}\right)  \right.
\\ &  \qquad -\frac{({\mathbb{I}}-3\widehat{\boldsymbol{x}} \widehat{\boldsymbol{x}})\mathrm{i}}{k||\mathbf{x}||}\left(\mathrm{e}^{-\mathrm{i}k||\mathbf{x}||}+\mathrm{e}^{\mathrm{i}k||\mathbf{x}||}\right)\left.-\frac{({\mathbb{I}}-3\widehat{\boldsymbol{x}} \widehat{\boldsymbol{x}})}{\left(k||\mathbf{x}||\right)^2}\left(\mathrm{e}^{-\mathrm{i}k||\mathbf{x}||}-\mathrm{e}^{\mathrm{i}k||\mathbf{x}||}\right)\right],
\end{split}\end{align}
with $\boldsymbol{x} \equiv ||\boldsymbol{x}||  \widehat{\boldsymbol{x}} $.

To compare our results to their classical counterpart, in a Green function approach, let us define a quantum Green function, by rewriting equation (\ref{eq:PsiBeta2}) in the form

\begin{equation} \label{eq:PsiBetabis}
\bm{\psi}^{A.p-multipolar}_\perp\left(\mathbf{x},t\right)=\int_{-\infty}^{+\infty}\mathrm{d}t' {\bf \overleftrightarrow{G}}^{A.p-multipolar}\left(\mathbf{x},t-t'\right)c_{\mathrm{e}}\left(t'\right)\mathrm{e}^{-\mathrm{i}\omega_0 t'}\bm{\mu}_{g-e.},
\end{equation} where  ${\bf \overleftrightarrow{G}}\left(\mathbf{x},t-t'\right)c_{\mathrm{e}}\left(t'\right)\mathrm{e}^{-\mathrm{i}\omega_0 t'}\bm{\mu}_{g-e.}$ can be seen to represent the quantum electric field generated, at time $t$ and location  $\mathbf{x}$ by a (complex) dipole located at the origin ($\mathbf{x}=0$), at time $t'$ and oriented along $\widehat{\bm{\mu}}_{g-e.}$. This dipole is characterised by a (complex) amplitude $c_{\mathrm{e}}\left(t'\right)\mathrm{e}^{-\mathrm{i}\omega_0 t'}$ for $t'\geq 0$ and 0 otherwise. We get thus

\begin{equation} {\bf \overleftrightarrow{G}}^{A.p-multipolar}\left(\mathbf{x},t-t'\right)=\Theta(t-t')
(-\mathrm{i})\frac{c\omega_0} {8\pi^2\epsilon_0} \int_0^{+\infty}\mathrm{d}kk^2\frac{\mathrm{e}^{-\mathrm{i}ck(t-t')}}{ \left[1+\left(\frac{2}{3}a_0||\mathbf{k}||\right)^{2}\right]^2} \mathbf{\overleftrightarrow{M}}\left(k,\mathbf{x}\right)\;,\label{GreenQ}
\end{equation} where we introduced the Heaviside function $\Theta$ to recall that by construction we do not take account of influences from the (classical) future ($t'>t$).

\subsection{Classical Green function (linear susceptibility) formalism.}
In the framework of Maxwell's theory, the Green function associated to the field radiated by a dipole is also well-known: in a linear susceptibility approach, one finds \cite{Jackson,Rahmani} that the electric field response to a dipolar excitation located at the origin of the spatial system of coordinates obeys (leaving aside a Dirac delta contribution which expresses the electric field generated by the classical dipole at its own location, see Supplementary Material, section 3, for more details): 

\begin{align}
&\overleftrightarrow{\boldsymbol{g}}^{class.}\left(  \boldsymbol{x},t-t'\right)
=\label{classG}\\ \nonumber
&-\frac{\left[  {\mathbb{I}}-\widehat{\boldsymbol{x}}\widehat{\boldsymbol{x}}\right]}{ 4\pi \varepsilon_0 ||\mathbf{x}|| c^2}
\delta^{(2)}(t-t'-||\mathbf{x}|| /c) -\frac{\left[  {\mathbb{I}}-3\widehat{\boldsymbol{x}}\widehat{\boldsymbol{x}}\right]}{ 4\pi \varepsilon_0 ||\mathbf{x}||^2 c}
\delta^{(1)}(t-t'-||\mathbf{x}|| /c) -\frac{\left[  {\mathbb{I}}-3\widehat{\boldsymbol{x}}\widehat{\boldsymbol{x}}\right]}{ 4\pi \varepsilon_0 ||\mathbf{x}||^3 }\delta^{(0)}(t-t'-||\mathbf{x}|| /c)
\end{align}
where 
$\int_{-\epsilon}^{+\epsilon}du \delta^{(i)}(u)f(u)=\left(-1\right)^i\frac{d^i}{du^i}f(u)$ estimated at $u=0$ (for $i=0,1,2$). The Green function (\ref{classG}) obviously respects causality, due to the presence of the causal delay $\delta t=t-t'=||\mathbf{x}|| /c$ \cite{Brill}.

As is also shown in Supplementary Material (section 3), we obtain, separating the transverse and longitudinal responses of the Green function (again leaving a Dirac-delta singularity at $\boldsymbol{x}=\bf{0}$ aside): \begin{align}\overleftrightarrow{\boldsymbol{g}}^{class.}\left(  \boldsymbol{x},t\right)=\overleftrightarrow{\boldsymbol{g}}^{class.}_{\parallel}\left(  \boldsymbol{x},t\right)+\overleftrightarrow{\boldsymbol{g}}^{class.}_{\perp}\left(  \boldsymbol{x},t\right)\end{align} with
 \begin{align}
\overleftrightarrow{\boldsymbol{g}}^{\rm class.}_{\parallel}\left(  \boldsymbol{x},t-t'\right)
&=-\frac{\left[  {\mathbb{I}}-3\widehat
{\boldsymbol{x}}\widehat{\boldsymbol{x}}\right] }{4\pi \varepsilon_0 ||\mathbf{x}||^{3} 
 }\delta^{(0)}(t-t') \label{longi}\end{align}
 and\begin{align}\begin{split}\label{crux1}   \overleftrightarrow{\boldsymbol{g}}^{\rm class.}_{\perp}\left(  \boldsymbol{x},t-t'\right) &=\frac{\left[  {\mathbb{I}}-3\widehat{\boldsymbol{x}}\widehat{\boldsymbol{x}}\right] }{4\pi \varepsilon_0 ||\mathbf{x}||^{3} }\delta^{(0)}(t-t') 
 -\frac{\left[  {\mathbb{I}}-\widehat{\boldsymbol{x}}\widehat{\boldsymbol{x}}\right]}{ 4\pi \varepsilon_0 ||\mathbf{x}|| c^2}\delta^{(2)}(t-t'-||\mathbf{x}|| /c) \\&  \qquad
-\frac{\left[  {\mathbb{I}}-3\widehat{\boldsymbol{x}}\widehat{\boldsymbol{x}}\right]}{ 4\pi \varepsilon_0 ||\mathbf{x}||^2 c}\delta^{(1)}(t-t'-||\mathbf{x}|| /c)
-\frac{\left[  {\mathbb{I}}-3\widehat{\boldsymbol{x}}\widehat{\boldsymbol{x}}\right]}{ 4\pi \varepsilon_0 ||\mathbf{x}||^3 }\delta^{(0)}(t-t'-||\mathbf{x}|| /c) \;.
\end{split}\end{align}
 
Both the transverse and the longitudinal Green functions possess in the near-field an instantaneous component
corresponding to a static dipole {\it \`a la} Coulomb, opposite in sign however, so that they exactly cancel each other, which explains why the full response (\ref{classG}) is still causal. 
\subsection{Comparison with classical results: E.r coupling at the dipolar approximation.}
Injecting into (\ref{GreenQ}) the (\rw{dipolar}) E.r coupling factor (\ref{eq:MatrixDipole}) instead of its (\rw{multipolar}) A.p counterpart (\ref{eq:MatrixMultipole}),  we obtain the ``standard'' quantum Green function (see Ref.\cite{Foxmulder}, complement 10A, for a similar derivation in the case of the far field, also in the dipole approximation)
\begin{equation} {\bf \overleftrightarrow{G}}^{E.r-dip.}\left(\mathbf{x},t-t'\right)=\Theta(t-t')(-\mathrm{i})\frac{c}{8\pi^2\epsilon_0}\int_0^{+\infty} \mathrm{d}k k^2   (ck)  \mathrm{e}^{-\mathrm{i}ck(t-t')} \mathbf{\overleftrightarrow{M}}\left(k,\mathbf{x}\right),\label{totoEr}
\end{equation}

Making use of 
\begin{eqnarray} \label{eq:DipoleGreenbis2}
& \mathbf{\overleftrightarrow{M}}\left(k,\mathbf{x}\right)=\\ \nonumber & \frac{\mathrm{i} }{k||\mathbf{x}||}
 \left[({\mathbb{I}}-\widehat{\boldsymbol{x}} \widehat{\boldsymbol{x}})\vphantom{\frac{1}{\left(k||\mathbf{x}||\right)^2}} \left(\mathrm{e}^{-
 \mathrm{i}k||\mathbf{x}||}-\mathrm{e}^{\mathrm{i}k||\mathbf{x}||}\right)   
 -\frac{({\mathbb{I}}-3\widehat{\boldsymbol{x}} \widehat{\boldsymbol{x}})\mathrm{i}}{k||\mathbf{x}||}\left(\mathrm{e}^{-\mathrm{i}k
 ||\mathbf{x}||}+\mathrm{e}^{\mathrm{i}k||\mathbf{x}||}\right)
   -\frac{({\mathbb{I}}-3\widehat{\boldsymbol{x}} 
 \widehat{\boldsymbol{x}})}{\left(k||\mathbf{x}||\right)^2}\left(\mathrm{e}^{-\mathrm{i}k||\mathbf{x}||}-\mathrm{e}
 ^{\mathrm{i}k||\mathbf{x}||}\right) \right],
\end{eqnarray}
and extending the domain of integration of $k$ to minus infinity\footnote{As was pointed out by Hegerfeldt \cite{HegerfeldtFermi}, the absence of negative frequencies from the emitted field prevents causal propagation. However we do not consider this effect as a fundamental violation of causality. We interpret it instead as a blurring of the light cone \cite{CausalMidFar,VincePhd22}.} and  making use of the identities $\int_{-\infty}^{+\infty}\mathrm{d}k\mathrm{e}^{\mathrm{i}ku}=2\pi \delta(u)$, $\int_{-\infty}^{+\infty}\mathrm{d}k ik\mathrm{e}^{\mathrm{i}ku}=2\pi  \delta^{(1)}(u)$ and $\int_{-\infty}^{+\infty}\mathrm{d}k (-k)^2\mathrm{e}^{\mathrm{i}ku}=2\pi \delta^{(2)}(u)$, we finally get

\begin{eqnarray}\label{stdG}
&{\bf \overleftrightarrow{G}}^{E.r-dip.}\left(\mathbf{x},t-t'\right)=\\ \nonumber&
( -\frac{\left[  {\mathbb{I}}-\widehat{\boldsymbol{x}}\widehat{\boldsymbol{x}}\right]}{ 4\pi \varepsilon_0 ||\mathbf{x}|| c^2}
\delta^{(2)}(t-t'-||\mathbf{x}|| /c) 
-\frac{\left[  {\mathbb{I}}-3\widehat{\boldsymbol{x}}\widehat{\boldsymbol{x}}\right]}{ 4\pi \varepsilon_0 ||\mathbf{x}||^2 c}
\delta^{(1)}(t-t'-||\mathbf{x}|| /c) 
-\frac{\left[  {\mathbb{I}}-3\widehat{\boldsymbol{x}}\widehat{\boldsymbol{x}}\right]}{ 4\pi \varepsilon_0 ||\mathbf{x}||^3 }\delta^{(0)}(t-t'-||\mathbf{x}|| /c))
\end{eqnarray}

Comparing (\ref{classG}) and (\ref{stdG}), it is seen that the quantum Green function and the classical Green function are essentially the same function at this level of approximation. 
 Remark that only outgoing waves of the type $\mathrm{e}^{+\mathrm{i}k||\mathbf{x}||}$ contribute to the Green function (through $ \mathbf{\overleftrightarrow{M}}$  (\ref{eq:DipoleGreenbis})), the contribution of ingoing waves $\mathrm{e}^{-\mathrm{i}k||\mathbf{x}||}$ being cancelled by the Heaviside function $\Theta(t-t')$. This is a very appealing result in the sense that it establishes that the quantum e.m. field remains confined inside the light cone, a result known since Fermi \cite{FermiCausal}. However it remains an approximated result. What is puzzling here is that the quantum Green function that we derive in absence of longitudinal field is essentially equivalent to its classical counterpart, which at the other side results from the joint contribution of transverse and longitudinal fields. This would suggest that the classical-quantum correspondence is not fully verified, a rather suspicious hypothesis, at the origin of the present work.
In the next chapter we shall consider what happens if we use the A.p expression of the coupling instead of its E.r counterpart.

\subsection{Quantum Green function in the A.p coupling at the dipolar approximation.}

The factor $1/ \left[1+\left(\frac{2}{3}a_0||\mathbf{k}||\right)^{2}\right]^2$ in (\ref{eq:MatrixMultipole}) can actually \cite{VincePhd22} be interpreted as a cut off at frequencies larger than $\nu_{cut-off}\approx a_0/c$.  In addition to Hegerfeldt's cut-off of negative frequencies, this cut-off actually washes away spatial frequencies larger than $1/a_0$. Its implications have already been studied by two of us in detail in the past, in the case of the mid and far fields \cite{CausalMidFar,VincePhd22}, for which we showed for instance that it results in a blurring of the light cone, exponentially decreasing in space over a distance of the order of the Bohr length $a_0$ \cite{CausalMidFar,VincePhd22}. To simplify the discussion we shall thus formally displace the cut-off frequency to infinity ($ \left[1+\left(\frac{2}{3}a_0||\mathbf{k}||\right)^{2}\right]^2 = 1$).  In other words, in the next sections, we shall compute the A.p coupling factors in the dipolar approximation  (\ref{appro}). Accordingly we shall also integrate over negative frequencies, which means that we are not interested here in describing a blurring of the light cone {\it \`a la} Hegerfeldt. We get then, instead of (\ref{GreenQ})

\begin{equation} {\bf \overleftrightarrow{G}}^{A.p-dip.}\left(\mathbf{x},t-t'\right)=\Theta(t-t')
(-\mathrm{i})\frac{c\omega_0}{8\pi^2\epsilon_0} \int_{-\infty}^{+\infty}\mathrm{d}k k^2\mathrm{e}^{-\mathrm{i}ck(t-t')} \mathbf{\overleftrightarrow{M}}\left(k,\mathbf{x}\right),\label{GreenQbis}
\end{equation}
which, in agreement with (\ref{toto}), differs from its E.r counterpart (\ref{totoEr}) by a factor $\omega_0/k.c$ in the integrand.

The mid and far field  contributions to the quantum Green function can be easily computed as in the ``standard'' situation, making use of the identities $\int_{-\infty}^{+\infty}\mathrm{d}k\mathrm{e}^{\mathrm{i}ku}=2\pi \delta^{(0)}(u)$, $\int_{-\infty}^{+\infty}\mathrm{d}k ik\mathrm{e}^{\mathrm{i}ku}=2\pi  \delta^{(1)}(u)$. However, in the present case (A.p coupling), the near field contribution to the quantum Green function is proportional to the integral $\int_{-\infty}^{+\infty}\mathrm{d}k k^2\mathrm{e}^{-\mathrm{i}ck(t-t')}\frac{({\mathbb{I}}-3\widehat{\boldsymbol{x}} \widehat{\boldsymbol{x}})}{\left(k||\mathbf{x}||\right)^3}\left(\mathrm{e}^{-\mathrm{i}k||\mathbf{x}||}-\mathrm{e}^{\mathrm{i}k||\mathbf{x}||}\right)$, singular at $k=0$, which must be tackled differently. Making use of the identity $\int_{-\infty}^{+\infty}\frac{\mathrm{d}k}{k}\mathrm{e}^{\mathrm{i}ku}=i\pi (\theta(u)-\theta(-u))$, it is easy to show that this integral is proportional to 

$i\pi(\theta(-(t-t')-||\mathbf{x}|| /c)-\theta((t-t')+||\mathbf{x}|| /c){\bf -}\theta(-(t-t')+||\mathbf{x}|| /c){\bf +}\theta((t-t')-||\mathbf{x}|| /c))$

=$i\pi(0-1-\theta(-(t-t')+||\mathbf{x}|| /c)+\theta((t-t')-||\mathbf{x}|| /c))$ = $-2i\pi\cdot \theta(-(t-t')+||\mathbf{x}|| /c)$. Therefore the near field also contains contributions which are entirely outside the causal past of $(\mathbf{x},t)$ (but inside its classical past $t'\leq t$) and we finally get 
\begin{align}\begin{split} \label{apG}  &{\bf \overleftrightarrow{G}^{\bf{A.p-dip}}}\left(\mathbf{x},t-t'\right)= i\Theta(t-t') \frac{\omega_0}{4\pi\epsilon_0}\frac{1}{c^2||\mathbf{x}||}\cdot \left(\left[  {\mathbb{I}}-\widehat{\boldsymbol{x}}\widehat{\boldsymbol{x}}\right]\delta^{(1)}(t-t'-||\mathbf{x}|| /c) \right. \\ & \ \left. +\frac{c}{||\mathbf{x}||}\left[  {\mathbb{I}}-3\widehat{\boldsymbol{x}}\widehat{\boldsymbol{x}}\right]\delta^{(0)}(t-t'-||\mathbf{x}|| /c) + \frac{c^2}{({||\mathbf{x}||})^2}\left[  {\mathbb{I}}-3\widehat{\boldsymbol{x}}\widehat{\boldsymbol{x}}\right](\delta^{(-1)}(t-t'-||\mathbf{x}|| /c)-\delta^{(-1)}(t-t'))\right)\;. \end{split}\end{align}

where $\int_{-\epsilon}^{+\epsilon}du \delta^{(-1)}(u)f(u)=\left(-1\right)Prim(f(u))$ estimated at $u=0$, with $Prim(f(u))$ the primitive of $f$.

\rw{Comparison with (\ref{stdG}) shows that the result features not only an extra causal contribution, but also a completely noncausal term. This is a feature of the slow dependence of the A.p atom-field coupling on the norm of the electromagnetic wave vector $\mathbf{k}$. As far as we know, similar calculations \cite{FermiCausal,Foxmulder,PassanteVirtual,Sipe} of the outgoing field have mostly been carried out in the Power-Zineau-Woolley picture \cite{CohenQED1} of quantum electrodynamics where the interaction Hamiltonian is of the form $\hat{\mathbf{E}}\cdot\hat{\mathbf{x}}$ (E.r coupling). In this case the integrand in the near field contribution to the quantum Green function features no singularity at $k=0$ and one retrieves a causal result, in the framework of Wigner-Weisskopf decay with the dipole approximation, and with the extra approximation of extending the range of integration over wave numbers from the positive real semi-axis to the whole real axis.}
\subsection{A.p coupling: Comparison with standard and classical results}
We now see that in the dipolar approximation the quantum Green function in the A.p coupling differs, \rw{in the mid and far fields}, from its classical counterpart (\ref{classG}), and also from the standard  quantum Green function in the E.r coupling (\ref{totoEr}) by global multiplicative constant factors, but also by a shift in the order of differentiation. Now, because the amplitude $c_e$ depends exponentially on time \cite{EdouardIsa}, according to the Wigner-Weisskopf prediction ($c_e(t)\mathrm{e}^{-\mathrm{i}\omega_0t}\approx \mathrm{e}^{-\mathrm{i}\Omega_0t}$ with $\Omega_0=\omega_0-i (\Gamma/2)$), the standard (E.r) and A.p electric mid and far fields are proportional to each other.

For instance, in the dipolar approximation, the single-photon far field obeys 

$\bm{\psi}^{\bf{E.r-dip.}far}=(\Omega_0^2)  \mathrm{e}^{-\mathrm{i}\Omega_0(t-t'-||\mathbf{x}|| /c)}$$\Theta(t-t')\frac{1}{4\pi\epsilon_0 c^2}\frac{1}{||\mathbf{x}||}\cdot \left[  {\mathbb{I}}-\widehat{\boldsymbol{x}}\widehat{\boldsymbol{x}}\right]\bm{\mu}_{g-e.}$ in the E.r coupling, while, in the A.p coupling, it reads

$\bm{\psi}^{\bf{A.p-dip.}far}=(i\omega_0)(-i\Omega_0 )   \mathrm{e}^{-\mathrm{i}\Omega_0(t-t'-||\mathbf{x}|| /c)}$ $\Theta(t-t')\frac{1}{4\pi\epsilon_0  c^2}\frac{1}{||\mathbf{x}||}\cdot \left[  {\mathbb{I}}-\widehat{\boldsymbol{x}}\widehat{\boldsymbol{x}}\right]\bm{\mu}_{g-e.}$

The standard and A.p electric mid and far fields are thus proportional to each other (up to a factor $(\omega_0/\Omega_0)$):

\begin{equation}\bm{\psi}^{\bf{A.p-dip.}mid-far}_\perp \left(\mathbf{x},t\right)=(\omega_0/\Omega_0)\bm{\psi}^{\bf{E.r-dip}.mid-far}_\perp\left(\mathbf{x},t\right)\end{equation}

In the weak coupling regime, $(\omega_0/\Omega_0)$ is very close to unity ($\omega_0/\Gamma \approx 10^{3}$ for the 1S-2P transition), which shows that for what concerns the mid and far fields, in the dipolar approximation, the classical (\ref{classG}), standard E.r (\ref{stdG}) and A.p (\ref{apG}) Green functions are essentially the same object. 

However this is is not so in the case of the near field, due to the presence of a near field contribution proportional to $\delta^{(-1)}(t-t')$. In order to properly estimate the near field contribution we firstly need to estimate the primitive $Prim(f(u))$ with $f$ proportional to $\mathrm{e}^{-\mathrm{i}\Omega_0t}$. This primitive is defined up to an additional constant which we fix by requiring that for short positive times the single photon electric field is close to zero, by continuity (integrating (\ref{eq:ExactEQg}) over time with $c_e$ continuous by pieces over time guarantees the continuity over time of the photon wave function).

This leads to $Prim(c_{\mathrm{e}}\left(t'\right)\mathrm{e}^{-\mathrm{i}\omega_0 t'}) =Prim(\mathrm{e}^{-\mathrm{i}\Omega_0t'})=\frac{i}{ \Omega_0}(\mathrm{e}^{-\mathrm{i}\Omega_0t'}-1)$

The (transversal) near field thus obeys (\ref{eq:PsiBetabis},\ref{apG})
$\bm{\psi}_\perp^{{\bf A.p-dip.} near}$

$=\frac{\omega_0}{4\pi\epsilon_0}\int dt' \left[  \frac{1}{ ({||\mathbf{x}||})^3}\left[  {\mathbb{I}}-3\widehat{\boldsymbol{x}}\widehat{\boldsymbol{x}}\right](\theta(t-t'-||\mathbf{x}|| /c)-\theta(t-t'))\right]\mathrm{e}^{-\mathrm{i}\Omega_0t'}\bm{\mu}_{eg}$, where the integration over $t'$ runs from $t'=t-||\mathbf{x}|| /c$ to $t'=t$ if we consider $ \left(\mathbf{x},t\right)$ inside the causal light cone (originating from the event corresponding to the sudden excitation of the atom $ \left(\mathbf{x}=\mathbf{0},t=0\right)$). It runs from $t'=0$ to $t'=t$ outside. Therefore 
 
 \begin{align} \label{eq:FinalTrans}
 \bm{\psi}_\perp^{{\bf A.p-dip.} near}&= \frac{1}{4\pi\epsilon_0} \frac{\omega_0}{\Omega_0}\left[  \frac{1}{({||\mathbf{x}||})^3}\left[  {\mathbb{I}}-3\widehat{\boldsymbol{x}}\widehat{\boldsymbol{x}}\right]\right]\bm{\mu}_{eg}\nonumber\\ &  \qquad .\left(\theta\left(-||\mathbf{x}||+ct\right)(\mathrm{e}^{-\mathrm{i}\Omega_0t}-\mathrm{e}^{-\mathrm{i}\Omega_0(t-||\mathbf{x}|| /c)})+ \theta\left(||\mathbf{x}||-ct\right)                      (\mathrm{e}^{-\mathrm{i}\Omega_0t}-1)\right),
 \end{align}
 which does not cancel outside the causal light cone. Note that these results perfectly match those obtained in the past making use of an alternative approach based on complex calculus and residues technique \cite{VincePhd22}. That alternative approach has the supplementary merit to take account of the factor $1/ \left[1+\left(\frac{2}{3}a_0||\mathbf{k}||\right)^{2}\right]^2$, which we assumed here to be equal to 1. As is shown in Ref.\cite{VincePhd22}, the blurring of the light cone due to the cut-off factor $1/ \left[1+\left(\frac{2}{3}a_0||\mathbf{k}||\right)^{2}\right]^2$ can be computed exactly, and it decouples from the strong violation of causality described here.

 \section{\rw{Restoring causality: analogy with classical Green functions.}\label{3}}
  By a direct comparison of (\ref{apG}) with (\ref{crux1}), we remark that the classical transverse Green function and the quantum ``transverse'' (photonic) Green function (in A.p coupling and in the dipolar approximation) are essentially equivalent, provided we identify the ``quantum source term'' with the primitive in time of the quantum dipole. Having in mind that in quantum optics the longitudinal field is not quantized but is on the contrary  treated classically \cite{Messiah2}, we conclude that, similar to what happens in Maxwell's theory, it is sufficient (and necessary) in order to restore causality to introduce an instantaneous longitudinal field (\ref{longi}) with the same source term as for the transverse field.
 This means that we must add to the transverse quantum field (always leaving aside a Dirac delta at the origin which is irrelevant for the discussion) an instantaneous quantum longitudinal field $\bm{\psi}_\parallel^{{\bf A.p-dip.} near}$ = $\frac{1}{4\pi\epsilon_0} \frac{\omega_0}{\Omega_0}\left[  \frac{1}{({||\mathbf{x}||})^3}\left[  {\mathbb{I}}-3\widehat{\boldsymbol{x}}\widehat{\boldsymbol{x}}\right]\right]\bm{\mu}_{eg}(1-\mathrm{e}^{-\mathrm{i}\Omega_0t})$ inside and outside the light cone.
After cancellation of the longitudinal and transverse contributions to the instantaneous near field (which only happens outside the light cone), we find that the resulting field is well confined inside the light cone, so that causality is respected:
 
 \begin{align}\begin{split}\bm{\psi}^{{\bf A.p-dip.} near}&=\bm{\psi}_\perp^{{\bf A.p-dip.} near}+\bm{\psi}_\parallel^{{\bf A.p-dip.} near}\\ &=\left[\theta\left(-||\mathbf{x}||+ct\right)(1-\mathrm{e}^{-\mathrm{i}\Omega_0(t-||\mathbf{x}|| /c)})\right] \frac{1}{4\pi\epsilon_0}\frac{\omega_0}{\Omega_0} \left[  \frac{1}{({||\mathbf{x}||})^3}\left[  {\mathbb{I}}-3\widehat{\boldsymbol{x}}\widehat{\boldsymbol{x}}\right]\bm{\mu}_{eg}\right].\label{final}\end{split}\end{align}

 A new feature appears at this level however: contrary to the mid and far field which are proportional to each other, the single photon electric near field is no longer proportional to $\mathrm{e}^{-\mathrm{i}\Omega_0(t-||\mathbf{x}|| /c)}$ but it is proportional to $\mathrm{e}^{-\mathrm{i}\Omega_0(t-||\mathbf{x}|| /c)}-1$, due to the presence of an additive integration constant (that we fixed by imposing the continuity of the transverse field for short times). Accordingly, we find now that 

\begin{align}\begin{split}\bm{\psi}^{{\bf A.p-dip.} near} \left(\mathbf{x},t\right) & =\frac{(1-\mathrm{e}^{-\mathrm{i}\Omega_0(t-||\mathbf{x}|| /c)})}{(-\mathrm{e}^{-\mathrm{i}\Omega_0(t-||\mathbf{x}|| /c)})}(\omega_0/\Omega_0)\bm{\psi}^{{\bf E.r-dip.} near}\left(\mathbf{x},t\right) \\ & \not=(\omega_0/\Omega_0)\bm{\psi}^{{\bf E.r-dip.} near}\left(\mathbf{x},t\right)\label{totodiff}\end{split}\end{align}

 Henceforth, we predict the appearance of a surprising effect which consists of a remnant static field inside the causal light cone equal to $\theta\left(-||\mathbf{x}||+ct\right) \frac{1}{4\pi\epsilon_0} \frac{\omega_0}{\Omega_0}\left[  \frac{1}{({||\mathbf{x}||})^3}\left[  {\mathbb{I}}-3\widehat{\boldsymbol{x}}\widehat{\boldsymbol{x}}\right]\bm{\mu}_{eg}\right]$, still present a long time after the atom returned to the ground state. We are forced to interpret this prediction as follows: the e.m. field surrounding the atom ''remembers'' that the atom was excited and emitted a photon in the past, and a static field survives in the near field domain, disturbing the e.m. environment of the atom, a priori forever, or at least until another excitation takes place...

\section{Alternative definition of the quantum longitudinal field} \label{4}

An alternative approach, put forward in Ref.~\cite{VincePhd22}, is to consider, taking example on Eq.~(\ref{eq:GeneralWaveFunction}), the same matrix element for the longitudinal part of the electric field: 
\begin{equation} \label{eq:PsiLongit}
\bm{\psi}_\parallel\left(\mathbf{x},t\right)\equiv\langle\mathrm{g}\left(t\right),0\!\mid\hat{\mathbf{E}}_\parallel\left(\mathbf{x},0\right)\mid\!\psi\left(t\right)\rangle.
\end{equation}
Hence we have introduced a longitudinal photon wave function in an explicit and consistent way, as proposed in Ref.~\cite{MulderLongit}. The longitudinal electric field operator is given \cite{CohenQED1} by
\begin{equation} \label{eq:ElectricChargeField}\hat{\mathbf{E}}_\parallel\left(\mathbf{x},t\right)=-\frac{1}{4\pi\epsilon_0}\nabla\int\mathrm{d}\mathbf{x}'\frac{\hat{\rho}\left(\mathbf{x}',t\right)}{\|\mathbf{x}-\mathbf{x}'\|}
\end{equation}
with $\hat{\rho}$ the charge density operator
\begin{equation} \label{eq:ChargeDensity}
\hat{\rho}\left(\mathbf{x},t\right)=\sum_\alpha q_\alpha\,\delta\left(\mathbf{x}-\hat{\mathbf{x}}_\alpha\left(t\right)\right),
\end{equation}
for which the matrix element is
\begin{align}\begin{split} \label{eq:CDDirect}
\langle\mathrm{g}\left(t\right),0\!\mid\hat{\rho}\left(\mathbf{x},0\right)\mid\!\psi\left(t\right)\rangle&=\langle\mathrm{g},0\!\mid\hat{\rho}\left(\mathbf{x},t\right)\mid\!\psi\left(t=0\right)\rangle \\
&=-e\,\mathrm{e}^{-\mathrm{i}\Omega_0 t}\,\psi_\mathrm{g}^*\left(\mathbf{x}\right)\psi_\mathrm{e}\left(\mathbf{x}\right).
\end{split}\end{align}
As can be seen from Eq.~(\ref{eq:CDDirect}), obtaining the longitudinal wave function thus requires the computation \cite{VincePhd22} of the Coulomb-type integral
\begin{equation} \label{eq:OverPrime}
\begin{aligned} [b]
&\int\mathrm{d}\mathbf{x}'\frac{\psi_{1\mathrm{s}}^*\left(\mathbf{x}'\right)\psi_{2\mathrm{p}\,m_2}\left(\mathbf{x}'\right)}{\left\vert\mathbf{x}-\mathbf{x}'\right\vert}=\int\mathrm{d}\mathbf{x}'\frac{\mathrm{e}^{-\frac{3}{2}\frac{\left\vert\mathbf{x}'\right\vert}{a_0}}}{\left\vert\mathbf{x}-\mathbf{x}'\right\vert}\frac{1}{2^{\frac{5}{2}}\pi a_0^4}\left(\mathbf{x}'\cdot\bm{\xi}_{m_2}\right)\\
&=\frac{\xi_{m_2}^{\left(z\right)}}{2^{\frac{3}{2}}a_0^4}\int_0^{+\infty}\mathrm{d}r\,r^3\,\mathrm{e}^{-\frac{3}{2}\frac{r}{a_0}}
\int_{-1}^1\mathrm{d}\eta\,\frac{\eta}{\sqrt{r^2+\mathbf{x}^2-2\eta r\left\vert\mathbf{x}\right\vert}}\\
&=-\frac{\xi_{m_2}^{\left(z\right)}}{2^{\frac{3}{2}}3a_0^4\mathbf{x}^2}\int_0^{+\infty}\mathrm{d}r\,r\,\mathrm{e}^{-\frac{3}{2}\frac{r}{a_0}}\left[\left(r+\left\vert\mathbf{x}\right\vert\right)\right.
\left.\times\left(-r^2-\mathbf{x}^2+r\left\vert\mathbf{x}\right\vert\right)+\left\vert r-\left\vert\mathbf{x}\right\vert\right\vert\left(r^2+\mathbf{x}^2+r\left\vert\mathbf{x}\right\vert\right)\right]\\
&=-\frac{2^{\frac{3}{2}}}{3^5}\xi_{m_2}^{\left(z\right)}\frac{\mathrm{e}^{-\frac{3}{2}\frac{\left\vert\mathbf{x}\right\vert}{a_0}}}{a_0}\times\left[27\frac{\left\vert\mathbf{x}\right\vert}{a_0}+72+96\frac{a_0}{\left\vert\mathbf{x}\right\vert}+64\frac{a_0^2}{\mathbf{x}^2}-64\mathrm{e}^{\frac{3}{2}\frac{\left\vert\mathbf{x}\right\vert}{a_0}}\frac{a_0^2}{\mathbf{x}^2}\right].
\end{aligned}
\end{equation}
Finally we have obtained
\begin{equation} \label{eq:FinalLong}
\bm{\psi}_\parallel\left(\mathbf{x},t\right)=-\frac{\mathrm{i}}{\omega_0}\frac{\mathrm{e}^{-\mathrm{i}\Omega_0t}}{4\pi\epsilon_0\left\vert\mathbf{x}\right\vert^3}\left[  {\mathbb{I}}-3\widehat{\boldsymbol{x}}\widehat{\boldsymbol{x}}\right]\cdot\bm{\mu}_{eg}+\bm{\psi}_\parallel^{\left(\mathrm{exp.\,dec.}\right)}\left(\mathbf{x},t\right)\end{equation}
where $\bm{\psi}_\parallel^{\left(\mathrm{exp.\,dec.}\right)}$ decays exponentially with increasing distance $\left\vert\mathbf{x}\right\vert$ from the atomic nucleus (this term is the longitudinal equivalent of the Shirokov term, which vanishes in the dipole approximation performed above for the transverse wave function  \cite{VincePhd22}). To obtain Eq.~(\ref{eq:FinalLong}), we have used the expression $\omega_0=\left(3/8\right)\left(\alpha c/a_0\right)$ for the transition frequency.\\

Let us now consider the total wave function $\bm{\psi}_\parallel\left(\mathbf{x},t\right)+\bm{\psi}_\perp\left(\mathbf{x},t\right)$ of the emitted photon (in the dipole approximation), and investigate the apparently noncausal terms. By adding Eqs.~(\ref{eq:FinalTrans}) and~(\ref{eq:FinalLong}) we obtain (taking $\Omega_0\simeq\omega_0$)
\begin{align}\begin{split} \label{eq:NoNeedtoAsk}
\bm{\psi}\left(\mathbf{x},t\right)& =-\mathrm{i}\frac{1}{4\pi\epsilon_0\left\vert\mathbf{x}\right\vert^3}\frac{1}{\omega_0}\left[  {\mathbb{I}}-3\widehat{\boldsymbol{x}}\widehat{\boldsymbol{x}}\right]\cdot\bm{\mu}_{eg}\\
& \qquad \times\left[\theta\left(ct-\left\vert\mathbf{x}\right\vert\right)\mathrm{e}^{-\mathrm{i}\frac{\Omega_0}{c}\left(ct-\left\vert\mathbf{x}\right\vert\right)}+\left(1-\theta\left(ct-\left\vert\mathbf{x}\right\vert\right)\right)\right]+\bm{\psi}_\parallel^{\left(\mathrm{exp.\,dec.}\right)}\left(\mathbf{x},t\right),
\end{split}\end{align}
to be compared to Eq.~(\ref{final}). The first summand between the square brackets is causal and propagates in spacetime with an expected phase. The much more surprising second summand \emph{only subsists outside the light cone}, which constitutes a very counter-intuitive result. It is worth noting, however, that this violation of causality remains limited to the near field and gradually disappears while the light cone regularly expands with time.

\section{Conclusions}
 \rw{The problem of computing the electromagnetic field emitted during an atomic transition from an excited electronic state to a less energetic one (called, for our purposes here, the ground state) has long been a recurring theme of investigation in atomic physics and quantum electrodyamics \cite{FermiCausal,Shirokov,PassanteVirtual,HegerfeldtFermi,Sipe}. The first well-known investigation was carried out by Fermi \cite{FermiCausal}, who found that, at the Wigner-Weisskopf aproximation \cite{WW}, the emitted electromagnetic field propagates causally. Shirokov \cite{Shirokov} then noticed that Fermi's result involved extending the range of integration over electromagnetic frequencies from the positive real semi-axis to the whole real axis, thereby including nonphysical electromagnetic negative frequency modes in the treatment. Two of us have commented \cite{DDDNVZHaroche} on the soundness of this approximation in a different but similar framework. Hegerfeldt \cite{HegerfeldtFermi} generalised Shirokov's remarks, linking the absence of negative electromagnetic frequencies with the noncausal field propagation via Paley and Wiener's theorem \cite{PaleyWiener}. As we have shown elsewhere, the cut-off at negative frequencies results in a blurring of the light cone \cite{CausalMidFar,VincePhd22}. This result holds in the dipolar approximation (in the A.p and the E.r coupling as well) but gets reinforced when multipolar contributions are taken into account due to the presence of an additional cut-off at high frequencies. In the A.p coupling, this cut-off results, at the level of the mid and far fields, into a blurring outside the light cone exponentially decreasing over a distance of the order of the Bohr radius \cite{CausalMidFar}. This blurring cannot thus be considered as a strong violation of causality however because the source itself possesses some spatial extent.}

 \rw{If we consider the near field however,} there appear substantial differences between the A.p expression (\ref{apG}) and its ``standard'' (E.r) counterpart (\ref{stdG}); both expressions differ by the presence of an instantaneous response to the source term\footnote{\rw{Expressions (\ref{apG},\ref{stdG}) were obtained aft the dipolar approximation but it is worth noting that even if we take account of multipolar corrections, the conclusions of our analysis remain unchanged, up to a blurring over a distance of the order of the Bohr radius, as has been shown in Ref.  \cite{VincePhd22} via an exact computation based on residues technique. }} which constitutes a ``strong'' violation of causality. (\ref{apG}) also differs from its classical counterpart (\ref{crux1}) (where such an instantaneous transverse field is nevertheless present) in the sense that the role of the source term appears here to be played by the primitive over time of the amplitude of the quantum dipole.

In order to restore causality a possible strategy consists (as done in section 3), in full analogy with the classical situation, in adding to the transverse near field the longitudinal near field (equivalent to the classical Coulomb dipolar electric field (\ref{longi}), derived classically, but with the same quantum source term as for the quantum transverse field). 

Another way to tackle the problem consists (as done in section 4) of adding to the transverse field the longitudinal field having as a source term $c_{\mathrm{e}}\left(t'\right)\mathrm{e}^{-\mathrm{i}\omega_0 t'}$ and not its primitive $\frac{i}{\Omega_0}(\mathrm{e}^{-\mathrm{i}\Omega_0t'}-1)$. In this alternative approach, the price to pay is that causality is violated, but the benefit is that no remanent near field survives after the emission process.

In both cases, we predict the cancellation of a part of the photonic transverse field by the longitudinal field which is never interpreted in terms of photons \cite{Messiah2,AspectGrangierbis,Aspect}. At least our derivations emphasise the necessary coexistence of supposedly classical (longitudinal) and quantum (transverse) degrees of freedom, which could be expressed through the following paradox: either the probability of having a click in a photodetector results from the interference of transverse and longitudinal fields, or causality is violated. This view contradicts an apparently widely accepted opinion according to which ...{\it ``only transverse fields can be self-perpetuating fields (i.e. radiation fields that keep existing even if their source disappears''... (quotation from Ref. \cite{Dutra})}. We hope that a fully relativistically covariant formulation of the problem {\it \`a la} Gupta-Bleuler \cite{SurajNGupta} could contribute to transcend this apparent paradox, but presently this is an open question.

In any case, our computations establish that the near field (\ref{final}) predicted in the quantum approach \rw{in the A.P coupling} differs both from its classical counterpart, \rw{and from the field predicted in the E.r coupling, an intriguing result}. Lamb, quoted in Ref.\cite{Foxmulder} pointed out {\it ...a difference in measurable quantities...} predicted in the E.r and in the A.p coupling and indeed, in the last resort, experiment ought to reveal the existence of the effects predicted here. \rw{Of course it is not easy to probe the near field of an individual atom \cite{KellerBook} but the existence of collective memory effects has already been put into evidence in the past \cite{Memory} and, although it is out of the scope of the present paper it could be worth trying to relate them to the non-standard properties of the atomic near field investigated here. }

\section*{Acknowledgements}
\rw{We thank Edouard Brainis, Margaret Hawton, Boris Gralak  and Andr\'{e} Nicolet for interesting discussions. T.D. acknowledges support from the COST 1403 action, and thanks N. Sandeau for drawing his attention on reference \cite{Memory}. V.D. acknowledges support from CNRS (INSIS doctoral grant).}

\bibliographystyle{unsrt}
\bibliography{Bibliobisv17}

\section*{Supplementary material 1. A.p versus E.r coupling at the dipolar approximation.}

Estimating the coupling factor $G_\lambda\left(\mathbf{k}\right)=\langle1\mathrm{s},1_{\lambda,\mathbf{k}}\mid\!\hat{H}_I\!\mid\!2\mathrm{p}\,m_2,0\rangle$ in the dipolar approximation for the E.r coupling (see e.g. sections 1.1., 5.1.2, 6.3 and 10A in Ref. \cite{Foxmulder}, having in mind that in that reference certain computations were performed over a finite volume $V$, finite volume that we formally took to be equal to 1 here, due to the fact that it actually disappears in the expression of Glauber's correlation function, after taking the limit where $V$ goes to infinity, which is a common procedure in QED), we find

\begin{equation} \langle1\mathrm{s},1_{\lambda,\mathbf{k}}\mid\!\hat{H}^{E.r-dip.}_I\!\mid\!2\mathrm{p}m_2,0\rangle=\langle1\mathrm{s}\mid\!\mathrm{-i}\sqrt{\frac{\hbar c  k}{16\pi^3\epsilon_0 }}\bm{\epsilon}_{\left(\lambda\right)}^*\left(\mathbf{k}\right)\cdot e\bm{r}\mid\!2 \mathrm{p}m_2\rangle,\end{equation}

where we  denoted $\bm{\epsilon}_{\left(\lambda\right)}$ the direction of the polarization of the e.m. mode, $a_0$ the Bohr radius, and $\omega_0$ the Bohr frequency ($\omega_0$=$\omega_e-\omega_g$).

If we evaluate the A.p coupling factor in the A.p approximation, we get instead\begin{eqnarray}\langle1\mathrm{s},1_{\lambda,\mathbf{k}}\mid\!\hat{H}^{A.p-dip.}_I\!\mid\!2\mathrm{p}\,m_2,0\rangle=\langle1\mathrm{s}\mid\!\sqrt{\frac{\hbar}{16\pi^3\epsilon_0 ck}}\bm{\epsilon}_{\left(\lambda\right)}^*\left(\mathbf{k}\right)\cdot \frac{e\hbar\mathrm{i}\mathbf{\nabla}}{m}\mid\!2 \mathrm{p}m_2\rangle\label{appro}\end{eqnarray}Making use of the constraints \begin{equation}\mathbf{E}=\frac{-\partial}{\partial t} \mathbf{A},\end{equation}valid in Coulomb gauge, and\begin{equation}\langle1\mathrm{s}\mid\![H_e^{hydrogen},\mathbf{r}]\mid\!2 \mathrm{p}m_2\rangle=-\hbar \omega_0\langle1\mathrm{s}\mid\!\mathbf{r}\mid\!2 \mathrm{p}m_2\rangle=i\hbar\langle1\mathrm{s}\mid\!\frac{\hbar\mathbf{\nabla}}{im}\mid\!2 \mathrm{p}m_2\rangle,\end{equation} where $H_e^{hydrogen}$ is the Hamiltonian of the electron inside the atom, which contains the kinetic energy and the Coulomb potential generated by the proton, we find \begin{equation}\langle1\mathrm{s},1_{\lambda,\mathbf{k}}\mid\!\hat{H}^{E.r-dip.}_I\!\mid\!2\mathrm{p}\,m_2,0\rangle/\langle1\mathrm{s},1_{\lambda,\mathbf{k}}\mid\!\hat{H}^{A.p-dip.}_I\!\mid\!2\mathrm{p}\,m_2,0\rangle=\frac{ck}{\omega_0}\label{toto},\end{equation}so that in the dipolar approximation, the A.p and E.r matrix elements differ by a dimensionless factor $ck/\omega_0$.

\section*{Supplementary material 2: Single photon wave function/electric field.}

The Glauber first order correlation  function reads \cite{GlauberKoh,Foxmulder}
\begin{equation} \label{eq:GeneralWaveFunction2}
\begin{aligned} [b]
\bm{\psi}_\perp\left(\mathbf{x},t\right)&\equiv\langle\mathrm{g},0\!\mid\hat{\mathbf{E}}_\perp\left(\mathbf{x},0\right)\mid\psi\left(t\right)\rangle\\
&=\sum_{\lambda=\pm}\int\mathrm{d}^3k\,\langle0\!\mid\hat{\mathbf{E}}_\perp\left(\mathbf{x},0\right)c_{\mathrm{g},\lambda}\left(\mathbf{k},t\right)\mathrm{e}^{-\mathrm{i}c||\mathbf{k}||t}\!\mid\!1_{\lambda,\mathbf{k}}\rangle\\
&=\mathrm{i}\sqrt{\frac{\hbar c}{2\epsilon_0}}\sum_{\lambda=\pm}\int \frac{\mathrm{d}^3 k}{ (2\pi)^3} \sqrt{||\mathbf{k}||}\mathrm{e}^{\mathrm{i}\left(\mathbf{k}\cdot\mathbf{x}-c||\mathbf{k}||t\right)}c_{\mathrm{g},\lambda}\left(\mathbf{k},t\right)\bm{\epsilon}_{\left(\lambda\right)}\left(\mathbf{k}\right)
\end{aligned}
\end{equation}
where $\hat{\mathbf{E}}_\perp$ represents the transverse part of the electric field operator defined through

\begin{equation} \label{eq:JustAMomentPos}
\mathbf{\hat{E}}_\perp\left(\mathbf{x},t\right)=\mathrm{i}\sqrt{\frac{\hbar c}{2\epsilon_0}}\sum_{\lambda=\pm}\int \frac{\mathrm{d}^3k}{(2\pi)^3} \sqrt{||\mathbf{k}||}\left[\hat{a}_{\left(\lambda\right)}\left(\mathbf{k},t\right)\bm{\epsilon}_{\left(\lambda\right)}\left(\mathbf{k}\right)\mathrm{e}^{\mathrm{i}\mathbf{k}\cdot\mathbf{x}}-\hat{a}_{\left(\lambda\right)}^\dagger\left(\mathbf{k},t\right)\bm{\epsilon}_{\left(\lambda\right)}^*\left(\mathbf{k}\right)\mathrm{e}^{-\mathrm{i}\mathbf{k}\cdot\mathbf{x}}\right]
\end{equation}
with the photon ladder operators obeying the commutation relations
\begin{equation} \label{eq:aaDagger}
\left[\hat{a}_{\left(\varkappa\right)}\left(\mathbf{k}\right),\hat{a}_{\left(\lambda\right)}^\dagger\left(\mathbf{q}\right)\right]=\delta\left(\mathbf{k}-\mathbf{q}\right)\delta_{\varkappa\lambda}.
\end{equation}

At this point, it comes in handy to notice that (\ref{eq:ExactEQg}) can be formally integrated, yielding
\begin{equation} \label{eq:FormalInt}
c_{\mathrm{g},\lambda}\left(\mathbf{k},t\right)=-\frac{\mathrm{i}}{\hbar}\int_0^t\mathrm{d}t'\,G_\lambda\left(\mathbf{k}\right)c_{\mathrm{e}}\left(t'\right)\mathrm{e}^{\mathrm{i}\left(c||\mathbf{k}||-\omega_{\mathrm{e}}+\omega_{\mathrm{g}}\right)t'}
\end{equation}
so that the single-photon wave function reads, in the most general case of our problem,
\begin{equation} \label{eq:FromEdouard}
\begin{aligned} [b]
&\bm{\psi}_\perp\left(\mathbf{x},t\right)=\sqrt{\frac{c}{\hbar\epsilon_0}}\sum_{\lambda=\pm}\int\tilde{\mathrm{d}k}||\mathbf{k}||\mathrm{e}^{\mathrm{i}\left(\mathbf{k}\cdot\mathbf{x}-c||\mathbf{k}||t\right)}\bm{\epsilon}_{\left(\lambda\right)}\left(\mathbf{k}\right)G_\lambda\left(\mathbf{k}\right)\int_0^t\mathrm{d}t'\,c_{\mathrm{e}}\left(t'\right)\mathrm{e}^{\mathrm{i}\left(c||\mathbf{k}||-\omega_{\mathrm{e}}+\omega_{\mathrm{g}}\right)t'}\\
&=-\mathrm{i}\frac{2^{\frac{9}{2}}}{3^4}\frac{\hbar e}{\epsilon_0m_ea_0}\sum_{\lambda=\pm}\int\tilde{\mathrm{d}k}||\mathbf{k}||\mathrm{e}^{\mathrm{i}\left(\mathbf{k}\cdot\mathbf{x}-c||\mathbf{k}||t\right)}\bm{\epsilon}_{\left(\lambda\right)}\left(\mathbf{k}\right)\frac{\bm{\epsilon}_{\left(\lambda\right)}^*\left(\mathbf{k}\right)\cdot\bm{\xi}_{m_2}}{\left[1+\left(\frac{||\mathbf{k}||}{k_\mathrm{X}}\right)^2\right]^2}\int_0^t\mathrm{d}t'\,c_{\mathrm{e}}\left(t'\right)\mathrm{e}^{\mathrm{i}\left(c||\mathbf{k}||-\omega_0\right)t'}\\
\end{aligned}
\end{equation}
where we introduced $\omega_0\equiv\omega_{\mathrm{e}}-\omega_{\mathrm{g}}$ and $k_{\mathrm{X}}\equiv3/\left(2a_0\right)$. The unit polarisation vectors obey the closure relation
\begin{equation} \label{eq:PolarSum}
\sum_{\lambda=\pm}\left(\epsilon_{\left(\lambda\right)}^i\right)^*\left(\mathbf{k}\right)\epsilon_{\left(\lambda\right)}^j\left(\mathbf{k}\right)=\delta^{ij}-\frac{k^ik^j}{\mathbf{k}^2}
\end{equation}
so that the wave function now is
\begin{eqnarray} \label{eq:ToMe}
\begin{aligned} [b]
&\bm{\psi}_\perp\left(\mathbf{x},t\right)=\\ \nonumber &-\mathrm{i}\frac{2^{\frac{9}{2}}}{3^4}\frac{\hbar e}{\epsilon_0m_ea_0}\int\tilde{\mathrm{d}k}||\mathbf{k}||\frac{\mathrm{e}^{\mathrm{i}\left(\mathbf{k}\cdot\mathbf{x}-c||\mathbf{k}||t\right)}}{\left[1+\left(\frac{||\mathbf{k}||}{k_\mathrm{X}}\right)^2\right]^2}\left(\bm{\xi}_{m_2}-\frac{\bm{\xi}_{m_2}\cdot\mathbf{k}}{\mathbf{k}^2}\,\mathbf{k}\right)\int_0^t\mathrm{d}t'\,c_{\mathrm{e}}\left(t'\right)\mathrm{e}^{\mathrm{i}\left(c||\mathbf{k}||-\omega_0\right)t'}.
\end{aligned}
\end{eqnarray}
Choosing a coordinate system for which $\mathbf{x}$ points along the third axis $\hat{z}$, we can compute the angular integrals:
\begin{align*}
&\mathbf{F}\left(k,||\mathbf{x}||\right)\equiv\int_0^\pi\mathrm{d}\theta\sin\theta\int_0^{2\pi}\mathrm{d}\varphi\left[\bm{\xi}_{m_2}-\left(\bm{\xi}_{m_2}\cdot\frac{\mathbf{k}}{||\mathbf{k}||}\right)\frac{\mathbf{k}}{||\mathbf{k}||}\right]\mathrm{e}^{\mathrm{i}k||\mathbf{x}||\cos\theta}\\
&=\int_0^\pi\mathrm{d}\theta\sin\theta\,\mathrm{e}^{\mathrm{i}k||\mathbf{x}||\cos\theta}\cdot \\ &\int_0^{2\pi}\mathrm{d}\varphi\left\{\left[\begin{array}{c}\xi_{m_2}^{\left(x\right)}\\\xi_{m_2}^{\left(y\right)}\\\xi_{m_2}^{\left(z\right)}\end{array}\right]-\left[\begin{array}{c}\sin\theta\cos\varphi\\\sin\theta\sin\varphi\\\cos\theta\end{array}\right]\left(\xi_{m_2}^{\left(x\right)}\sin\theta\cos\varphi+\xi_{m_2}^{\left(y\right)}\sin\theta\sin\varphi+\xi_{m_2}^{\left(z\right)}\cos\theta\right)\vphantom{\left[\begin{array}{c}\xi_{m_2}^{\left(1\right)}\\\xi_{m_2}^{\left(2\right)}\\\xi_{m_2}^{\left(z\right)}\end{array}\right]}\right\}\\
&=2\pi\int_0^\pi\mathrm{d}\theta\sin\theta\,\mathrm{e}^{\mathrm{i}k||\mathbf{x}||\cos\theta}\left[\begin{array}{c}\xi_{m_2}^{\left(x\right)}\left(1-\frac{1}{2}\sin^2\theta\right)\\\\\xi_{m_2}^{\left(y\right)}\left(1-\frac{1}{2}\sin^2\theta\right)\\\\\xi_{m_2}^{\left(z\right)}\sin^2\theta\end{array}\right]\\
&\equiv2\pi\,\mathbf{I}\left(k,||\mathbf{x}||\right).
\end{align*}
The integrals over $\theta$ give
\begin{eqnarray} 
&I^{\left(x,y\right)}\left(k,||\mathbf{x}||\right)=\mathrm{i}\frac{\xi_{m_2}^{\left(x,y\right)}}{k||\mathbf{x}||}\left[\vphantom{\frac{1}{\left(k||\mathbf{x}||\right)^2}}\left(\mathrm{e}^{-\mathrm{i}k||\mathbf{x}||}-\mathrm{e}^{\mathrm{i}k||\mathbf{x}||}\right)-\frac{\mathrm{i}}{k||\mathbf{x}||}\left(\mathrm{e}^{-\mathrm{i}k||\mathbf{x}||}+\mathrm{e}^{\mathrm{i}k||\mathbf{x}||}\right)-\frac{1}{\left(k||\mathbf{x}||\right)^2}\left(\mathrm{e}^{-\mathrm{i}k||\mathbf{x}||}-\mathrm{e}^{\mathrm{i}k||\mathbf{x}||}\right)\right],\nonumber\\ 
&I^{\left(z\right)}\left(k||\mathbf{x}||\right)=-2\mathrm{i}\frac{\xi_{m_2}^{\left(z\right)}}{k||\mathbf{x}||}\left[\vphantom{\frac{1}{\left(k||\mathbf{x}||\right)^2}}-\frac{\mathrm{i}}{k||\mathbf{x}||}\left(\mathrm{e}^{-\mathrm{i}k||\mathbf{x}||}+\mathrm{e}^{\mathrm{i}k ||\mathbf{x}||}\right)-\frac{1}{\left(k||\mathbf{x}||\right)^2}\left(\mathrm{e}^{-\mathrm{i}k||\mathbf{x}||}-\mathrm{e}^{\mathrm{i}k||\mathbf{x}||}\right)\right]\label{eq:DipoleGreen2}
\end{eqnarray}
As we can see, the photon wave function contains contributions proportional to $1/||\mathbf{x}||$, $1/||\mathbf{x}||^2$ and $1/||\mathbf{x}||^3$, which are respectively known as the far-field, mid-field and near-field contributions. We then have
\begin{equation} \label{eq:PsiBeta2bis}
\bm{\psi}_\perp\left(\mathbf{x},t\right)=-\mathrm{i}\frac{2^{\frac{7}{2}}}{3^4}\frac{\hbar e}{\epsilon_0m_ea_0}\int_0^{+\infty}\frac{\mathrm{d}k}{\left(2\pi\right)^2}k^2\frac{\mathrm{e}^{-\mathrm{i}ckt}}{\left[1+\left(\frac{k}{k_\mathrm{X}}\right)^2\right]^2}\mathbf{I}\left(k,||\mathbf{x}||\right)\int_0^t\mathrm{d}t'\,c_{\mathrm{e}}\left(t'\right)\mathrm{e}^{\mathrm{i}\left(ck-\omega_0\right)t'}.
\end{equation}
In most of what follows we will use the Wigner-Weisskopf approximation of exponential decay. We shall then have
\begin{equation}
c_{\mathrm{e}}\left(t\right)=\mathrm{e}^{-\frac{1}{2}\Gamma t}
\end{equation}
where  $\Gamma$ is the decay rate and $\omega_0$ the Bohr frequency of transition between the 1S and 2P states\footnote{Note that we neglect here the partial Lamb shift of the excited $2\mathrm{p}$ level due to the $1\mathrm{s}$ level, because it is a very small effect.}. This yields
\begin{equation} \label{eq:WWInt2}
\begin{aligned} [b]
&\int_0^t\mathrm{d}t'\,c_{\mathrm{e}}\left(t'\right)\mathrm{e}^{\mathrm{i}\left(ck-\omega_0\right)t'}=\int_0^t\mathrm{d}t'\,\mathrm{e}^{\mathrm{i}\left(ck-\omega_0\right)t'}\,\mathrm{e}^{-\frac{1}{2}\Gamma t'}\\
&=\left[\frac{\mathrm{e}^{\mathrm{i}\left(ck-\omega_0\right)t'}\,\mathrm{e}^{-\frac{1}{2}\Gamma t'}}{\mathrm{i}\left(ck-\omega_0\right)-\frac{1}{2}\Gamma}\right]_0^t=\mathrm{i}\,\frac{1-\mathrm{e}^{\mathrm{i}\left(ck-\omega_0\right)t}\,\mathrm{e}^{-\frac{1}{2}\Gamma t}}{ck-\omega_0+\frac{\mathrm{i}}{2}\Gamma}.
\end{aligned}
\end{equation}
We introduce the space-saving notation,
\begin{equation}
\Omega_0\equiv\omega_0-\frac{\mathrm{i}}{2}\Gamma,
\end{equation}
We can then write the contributions to the far-field, mid-field, and near-field to this photon wave function:
\begin{subequations} \label{eq:GiveMeaMoment}
\begin{eqnarray} \label{eq:NobodyCares}\nonumber
\psi_{\perp\left(\mathrm{far}\right)}^{\left(x,y\right)}&=\mathrm{i}\frac{2^{\frac{7}{2}}}{3^4}\frac{\hbar e}{\epsilon_0m_ea_0}\frac{\xi_{m_2}^{\left(x,y\right)}}{||\mathbf{x}||}\int_0^{+\infty}\frac{\mathrm{d}k}{\left(2\pi\right)^2}k\frac{\mathrm{e}^{-\mathrm{i}\Omega_0t}}{\left[1+\left(\frac{k}{k_\mathrm{X}}\right)^2\right]^2}\left(\mathrm{e}^{-\mathrm{i}k||\mathbf{x}||}-\mathrm{e}^{\mathrm{i}k||\mathbf{x}||}\right)\frac{1-\mathrm{e}^{-\mathrm{i}\left(ck-\Omega_0\right)t}}{ck-\Omega_0},\\ \nonumber
\psi_{\perp\left(\mathrm{mid}\right)}^{\left(x,y\right)}&=\frac{2^{\frac{7}{2}}}{3^4}\frac{\hbar e}{\epsilon_0m_ea_0}\frac{\xi_{m_2}^{\left(x,y\right)}}{||\mathbf{x}||^2}\int_0^{+\infty}\frac{\mathrm{d}k}{\left(2\pi\right)^2}\frac{\mathrm{e}^{-\mathrm{i}\Omega_0t}}{\left[1+\left(\frac{k}{k_\mathrm{X}}\right)^2\right]^2}\left(\mathrm{e}^{-\mathrm{i}k||\mathbf{x}||}+\mathrm{e}^{\mathrm{i}k||\mathbf{x}||}\right)\frac{1-\mathrm{e}^{-\mathrm{i}\left(ck-\Omega_0\right)t}}{ck-\Omega_0},\\ \nonumber
\psi_{\perp\left(\mathrm{near}\right)}^{\left(x,y\right)}&=-\mathrm{i}\frac{2^{\frac{7}{2}}}{3^4}\frac{\hbar e}{\epsilon_0m_ea_0}\frac{\xi_{m_2}^{\left(x,y\right)}}{||\mathbf{x}||^3}\int_0^{+\infty}\frac{\mathrm{d}k}{\left(2\pi\right)^2}\frac{1}{k}\frac{\mathrm{e}^{-\mathrm{i}\Omega_0t}}{\left[1+\left(\frac{k}{k_\mathrm{X}}\right)^2\right]^2}\left(\mathrm{e}^{-\mathrm{i}k||\mathbf{x}||}-\mathrm{e}^{\mathrm{i}k||\mathbf{x}||}\right)\frac{1-\mathrm{e}^{-\mathrm{i}\left(ck-\Omega_0\right)t}}{ck-\Omega_0},\\
\psi_{\perp\left(\mathrm{far}\right)}^{\left(z\right)}&=0,\nonumber\\ \nonumber
\psi_{\perp\left(\mathrm{mid}\right)}^{\left(z\right)}&=-\frac{2^{\frac{9}{2}}}{3^4}\frac{\hbar e}{\epsilon_0m_ea_0}\frac{\xi_{m_2}^{\left(z\right)}}{||\mathbf{x}||^2}\int_0^{+\infty}\frac{\mathrm{d}k}{\left(2\pi\right)^2}\frac{\mathrm{e}^{-\mathrm{i}\Omega_0t}}{\left[1+\left(\frac{k}{k_\mathrm{X}}\right)^2\right]^2}\left(\mathrm{e}^{-\mathrm{i}k||\mathbf{x}||}+\mathrm{e}^{\mathrm{i}k||\mathbf{x}||}\right)\frac{1-\mathrm{e}^{-\mathrm{i}\left(ck-\Omega_0\right)t}}{ck-\Omega_0},\\ \nonumber
\psi_{\perp\left(\mathrm{near}\right)}^{\left(z\right)}&=\mathrm{i}\frac{2^{\frac{9}{2}}}{3^4}\frac{\hbar e}{\epsilon_0m_ea_0}\frac{\xi_{m_2}^{\left(z\right)}}{||\mathbf{x}||^3}\int_0^{+\infty}\frac{\mathrm{d}k}{\left(2\pi\right)^2}\frac{1}{k}\frac{\mathrm{e}^{-\mathrm{i}\Omega_0t}}{\left[1+\left(\frac{k}{k_\mathrm{X}}\right)^2\right]^2}\left(\mathrm{e}^{-\mathrm{i}k||\mathbf{x}||}-\mathrm{e}^{\mathrm{i}k||\mathbf{x}||}\right)\frac{1-\mathrm{e}^{-\mathrm{i}\left(ck-\Omega_0\right)t}}{ck-\Omega_0}. \nonumber
\end{eqnarray}
\end{subequations}


\section*{Supplementary material 3: Classical Green function for dipole emission.} 

In the framework of Maxwell's theory, one finds \cite{Jackson,Rahmani} that the electric field response to a dipolar excitation located at the origin of the spatial system of coordinates obeys (in the harmonic regime)

\begin{eqnarray}\nonumber\overleftrightarrow{\boldsymbol{g}}^{class.}\left(  \boldsymbol{x},\omega\right)=-\frac{e^{ik||\mathbf{x}||}}{ 4\pi \varepsilon_0||\mathbf{x}||^{3}}\left\{  \left(1-ik||\mathbf{x}||-||\mathbf{x}||^{2}k^{2}\right)  \left[  {\mathbb{I}}-\widehat{\boldsymbol{x}}\widehat{\boldsymbol{x}}\right]   -\left(  2-2ik||\mathbf{x}||\right)  \widehat{\boldsymbol{x}}\widehat{\boldsymbol{x}}\right\}  - \frac{\mathbb{I} \delta^{3}\left(  \boldsymbol{x}\right) }{ 3\varepsilon_0}, \label{greluche}\end{eqnarray} with $\omega/k=c$.

Taking the temporal Fourier transform of (\ref{greluche}), and leaving aside the Dirac delta contribution which expresses the electric field generated by the classical dipole at its own location, we get the linear susceptibility in function of space and time:

\begin{eqnarray}
&&\overleftrightarrow{\boldsymbol{g}}^{class.}\left(  \boldsymbol{x},t-t'\right)\label{classG33}
=\\ \nonumber
&&-\frac{\left[  {\mathbb{I}}-\widehat{\boldsymbol{x}}\widehat{\boldsymbol{x}}\right]}{ 4\pi \varepsilon_0 ||\mathbf{x}|| c^2}\delta^{(2)}(t-t'-||\mathbf{x}|| /c)
-\frac{\left[  {\mathbb{I}}-3\widehat{\boldsymbol{x}}\widehat{\boldsymbol{x}}\right]}{ 4\pi \varepsilon_0 ||\mathbf{x}||^2 c}\delta^{(1)}(t-t'-||\mathbf{x}|| /c) 
-\frac{\left[  {\mathbb{I}}-3\widehat{\boldsymbol{x}}\widehat{\boldsymbol{x}}\right]}{ 4\pi \varepsilon_0 ||\mathbf{x}||^3 }\delta^{(0)}(t-t'-||\mathbf{x}|| /c)\end{eqnarray}

where 
$\int_{-\epsilon}^{+\epsilon}du \delta^{(i)}(u)f(u)=\left(-1\right)^i\frac{d^i}{du^i}f(u)$ estimated at $u=0$ (for $i=0,1,2$). The Green function (\ref{classG33}) obviously respects causality, due to the presence of the causal delay $\delta t=t-t'=||\mathbf{x}|| /c$ \cite{Brill}.

In order to separate the transverse and longitudinal responses of the Green function, it is necessary to go to momentum space where, making use  \cite{CohenQED1} of the projectors $\left[  {\mathbb{I}}-\widehat{\boldsymbol{k}}\widehat{\boldsymbol{k}}\right]$ and $  \widehat{\boldsymbol{k}}\widehat{\boldsymbol{k}}$ which respectively project plane waves onto their transverse and longitudinal components: $\overleftrightarrow{\boldsymbol{g}}^{class.}\left(  \boldsymbol{k},\omega\right)$=$\overleftrightarrow{\boldsymbol{g}}^{class.}_\perp\left(  \boldsymbol{k},\omega\right)\left[  {\mathbb{I}}-\widehat{\boldsymbol{k}}\widehat{\boldsymbol{k}}\right]$
+$\overleftrightarrow{\boldsymbol{g}}^{class.}_{\parallel}\left(  \boldsymbol{k},\omega\right)  \widehat{\boldsymbol{k}}\widehat{\boldsymbol{k}}$. By doing so, one finds \cite{Jackson,Rahmani} that the linear susceptibility  (in harmonic response), expressed in direct space through (\ref{greluche}) obeys the  decomposition $\overleftrightarrow{\boldsymbol{g}}^{class.}\left(  \boldsymbol{x},\omega\right)=\overleftrightarrow{\boldsymbol{g}}^{class.}_{\parallel}\left(  \boldsymbol{x},\omega\right)+\overleftrightarrow{\boldsymbol{g}}^{class.}_{\perp}\left(  \boldsymbol{x},\omega\right)$ with

\begin{equation}\nonumber
\overleftrightarrow{\boldsymbol{g}}^{class.}_{\parallel}\left(  \boldsymbol{x},\omega\right)
=-\frac{1}{4\pi \varepsilon_0 ||\mathbf{x}||^{3}}\left[  {\mathbb{I}}-3\widehat
{\boldsymbol{x}}\widehat{\boldsymbol{x}}\right]  - \frac{\mathbb{I} \delta^{3}\left(  \boldsymbol{x}\right) }{ 3\varepsilon_0} ; \label{glong2}
\end{equation}

\begin{eqnarray}\nonumber
\overleftrightarrow{\boldsymbol{g}}^{class.}_{\perp}\left(  \boldsymbol{x},\omega\right)
&=\frac{\left[  {\mathbb{I}}-3\widehat
{\boldsymbol{x}}\widehat{\boldsymbol{x}}\right] }{4\pi \varepsilon_0 ||\mathbf{x}||^{3} 
 }-\frac{e^{ik ||\mathbf{x}||}}{ 4\pi \varepsilon_0 ||\mathbf{x}||^{3}   }
 \left\{  \left(1-ik||\mathbf{x}||-||\mathbf{x}||^{2}k^{2}\right)  \left[  {\mathbb{I}}
-\widehat{\boldsymbol{x}}\widehat{\boldsymbol{x}}\right]   -\left(  2-2ik||\mathbf{x}||\right)  \widehat
{\boldsymbol{x}}\widehat{\boldsymbol{x}}\right\}.\label{blong}
\end{eqnarray}
In direct space, and in function of time, we get (again leaving the Dirac-delta singularity at $\boldsymbol{x}=\bf{0}$ aside),
 \begin{eqnarray}
\overleftrightarrow{\boldsymbol{g}}^{class.}_{\parallel}\left(  \boldsymbol{x},t-t'\right)
&=-\frac{\left[  {\mathbb{I}}-3\widehat
{\boldsymbol{x}}\widehat{\boldsymbol{x}}\right] }{4\pi \varepsilon_0 ||\mathbf{x}||^{3} 
 }\delta^{(0)}(t-t');\nonumber
\end{eqnarray}

 \begin{eqnarray}\nonumber
\overleftrightarrow{\boldsymbol{g}}^{class.}_{\perp}\left(  \boldsymbol{x},t-t'\right)
=\frac{\left[  {\mathbb{I}}-3\widehat
{\boldsymbol{x}}\widehat{\boldsymbol{x}}\right] }{4\pi \varepsilon_0 ||\mathbf{x}||^{3} 
 }\delta^{(0)}(t-t')+\overleftrightarrow{\boldsymbol{g}}^{class.}\left(  \boldsymbol{x},t-t'\right)=\frac{\left[  {\mathbb{I}}-3\widehat
{\boldsymbol{x}}\widehat{\boldsymbol{x}}\right] }{4\pi \varepsilon_0 ||\mathbf{x}||^{3} 
 }\delta^{(0)}(t-t') \\-\frac{\left[  {\mathbb{I}}-\widehat{\boldsymbol{x}}\widehat{\boldsymbol{x}}\right]}{ 4\pi \varepsilon_0 ||\mathbf{x}|| c^2}\delta^{(2)}(t-t'-||\mathbf{x}|| /c)
-\frac{\left[  {\mathbb{I}}-3\widehat{\boldsymbol{x}}\widehat{\boldsymbol{x}}\right]}{ 4\pi \varepsilon_0 ||\mathbf{x}||^2 c}\delta^{(1)}(t-t'-||\mathbf{x}|| /c) 
-\frac{\left[  {\mathbb{I}}-3\widehat{\boldsymbol{x}}\widehat{\boldsymbol{x}}\right]}{ 4\pi \varepsilon_0 ||\mathbf{x}||^3 }\delta^{(0)}(t-t'-||\mathbf{x}|| /c)\label{crux}.\nonumber
\end{eqnarray}


\end{document}